\documentclass[twocolumn,aps,prl]{revtex4}

\pdfoutput=1
\usepackage[pdftex]{graphicx}
\usepackage[unicode=true, bookmarks=false,bookmarksnumbered=true,bookmarksopen=true, breaklinks=false,backref=false,pagebackref=false, colorlinks=true]{hyperref}
\hypersetup{pdftitle={ },pdfauthor={ },linkcolor=black,citecolor=black, urlcolor=black, filecolor=black,pdfpagelayout=OneColumn,pdfnewwindow=true,pdfstartview=XYZ, plainpages=false}
\usepackage{rotating}
\usepackage{color}
\usepackage{amsmath}
\usepackage{braket}
\usepackage[T1]{fontenc}
\usepackage{url}
\usepackage[english]{babel}
\usepackage{dcolumn}
\usepackage{bm}
\usepackage{subfig}
\usepackage{float}
\usepackage{url}

\newcommand\T{\rule{0pt}{2.6ex}}
\newcommand\B{\rule[-1.2ex]{0pt}{0pt}}
\newcommand{\be}{\begin{equation}}
\newcommand{\ee}{\end{equation}}
\newcommand{\bea}{\begin{eqnarray}}
\newcommand{\eea}{\end{eqnarray}}

\newcommand{\bit}{\begin{itemize}}
\newcommand{\eit}{\end{itemize}}
\newcommand{\mbf}{\mathbf}

\graphicspath{{figures/}}
\usepackage{verbatim}

\newcommand{\inclfigs}

\begin{document}
\title{Colour Fields of the Static Pentaquark System\\ Computed in SU(3) Lattice QCD}

\author{Nuno Cardoso}
\email{nuno.cardoso@ist.utl.pt}
\author{Pedro Bicudo}
\email{bicudo@ist.utl.pt}
\affiliation{CFTP, Departamento de F\'{i}sica, Instituto Superior T\'{e}cnico, Universidade T\'{e}cnica de Lisboa, Av. Rovisco Pais, 1049-001 Lisbon, Portugal}

\begin{abstract}
We compute the colour fields of SU(3) lattice QCD created by static pentaquark systems, in a $24^3\times 48$
lattice at $\beta=6.2$
corresponding to a lattice spacing $a=0.07261(85)$ fm.
We find that the pentaquark colour fields are well described by a  multi-Y-type shaped flux tube.
The flux tube junction points are compatible with Fermat-Steiner points minimizing the total flux tube length.
We also compare the pentaquark flux tube profile with diquark-diantiquark central flux tube profile in the tetraquark and the quark-antiquark fundamental flux tube profile in the meson, and they match, thus showing that the pentaquark flux tubes are composed of fundamental flux tubes. 
\end{abstract}
\maketitle

\section{Introduction}

Here we study the colour field flux tubes produced by static pentaquarks in SU(3) lattice QCD. Unlike the colour fields of simpler few-body systems, say mesons, baryons and hybrids,
\cite{Ichie:2002dy,Okiharu:2004tg,Cardoso:2009kz,Cardoso:2010kw},
the pentaquark fields have not been previously studied in lattice QCD. 
This study is relevant both for the solution of theoretical problems and for the development of phenomenological models of QCD.

Quark confinement remains one of the main open theoretical problems of particle physics. In lattice QCD, flux tubes composed of colour-electric and colour-magnetic fields have been observed and this constitutes a very important clue for the understanding of quark confinement. Since the onset of QCD with its asymptotic freedom and infrared slavery, it is well known that confinement is due to the gluon fields and suppressed by the quark fields. It is thus important to measure the different possible flux tubes of pure gauge lattice QCD, to provide data for any theoretical attempt to solve the QCD confinement problem.

\begin{figure}[t!]
\begin{centering}
\includegraphics[width=8cm]{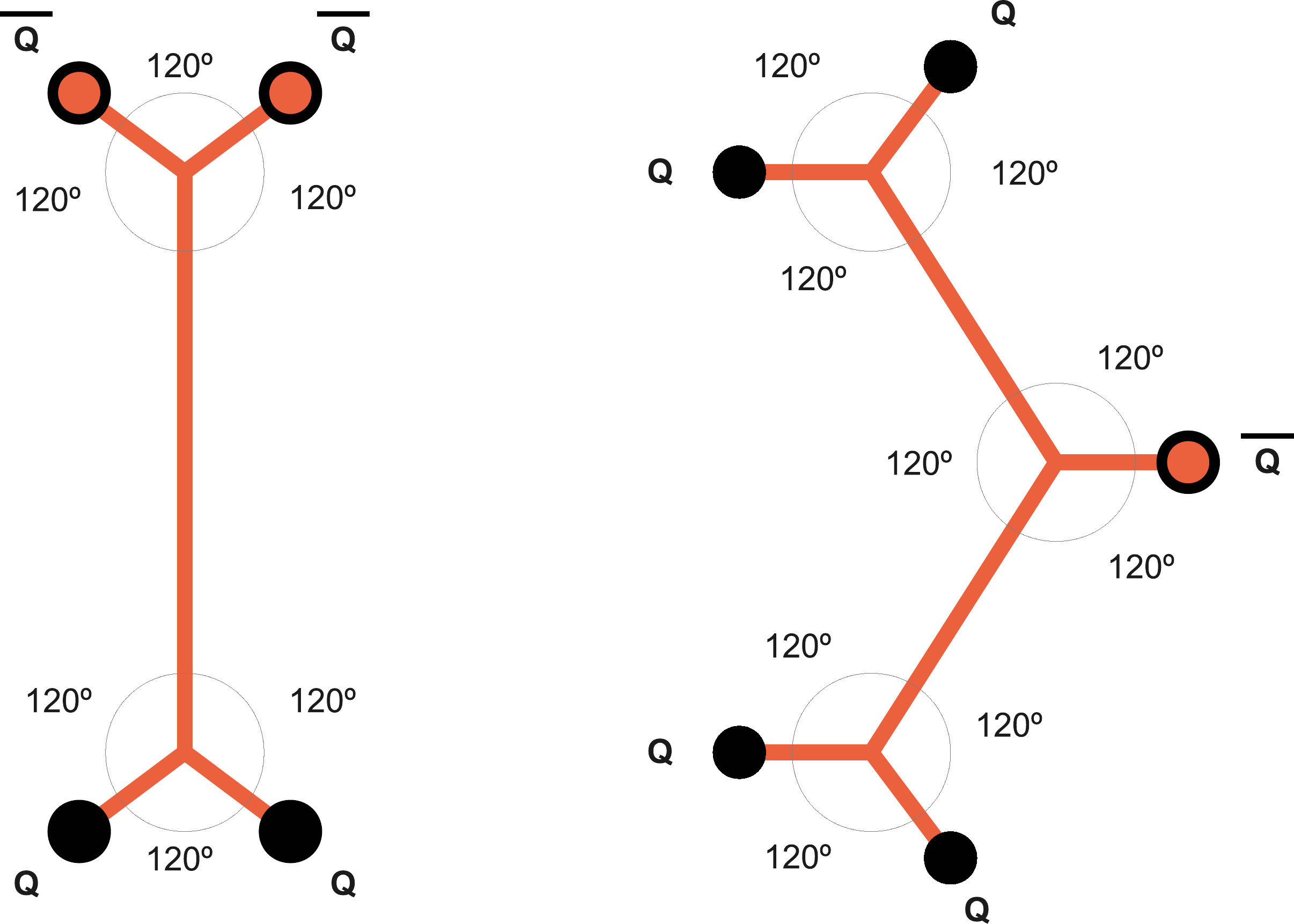}
\par\end{centering}
    \caption{(Colour online.)
In the string flip-flop model, thin elementary flux tubes similar to strings connect the colour charges in order to minimize the total length of the strings. Whenever geometrically possible, three elementary flux tubes meet  in a Fermat-Steiner point at an angle of  $\alpha=120^{\circ}$.  Here we depict planar examples of a tetraquark and a pentaquark flux tubes.}
    \label{fig:tq}
\end{figure}

Moreover, in what concerns phenomenology, the study of the colour fields in a pentaquark is important to discriminate between different multi-quark Hamiltonian models, quark models with two-body interactions only \cite{General:2007bk} as in the original quark model, from the string flip-flop model with a multi-body potential \cite{Bicudo:2010mv}. 
In the string flip-flop model, the colour charges are connected by strings disposed geometrically in order to minimize the total string length. The strings constitute the limit of very thin elementary flux tubes. An elementary or fundamental flux tube is the flux tube connecting the quark and antiquark of a meson, where the quark is in the triplet or fundamental representation of QCD. For instance in the two quark two antiquark system, depending of the position of these colour charges, the minimal string may be a two-meson string, or a tetraquark string, shaped like a double-Y flux tube, as in Fig. \ref{fig:tq},
composed of five linear fundamental flux tubes meeting in two Fermat-Steiner points \cite{Vijande:2007ix,Bicudo:2008yr,Richard:2009jv}.  A Fermat, or Steiner, point is defined as a junction minimizing the total length of strings, where linear individual strings join at $120^{\circ}$ angles. When the positions of the colour charges change, the potential may thus flip from one four-body potential to a pair of two-body potentials and flop back again. 
Notice the flip flop potential, compatible with the confining component of the flux tubes explored here, lead to  tetraquark boundstates,  below the strong decay threshold to pairs of mesons
\cite{Beinker:1995qe,Zouzou:1986qh,Gelman:2002wf,Vijande:2007ix}.
Recent investigations found that, even above the strong decay threshold, 
the presence of a centrifugal barrier in high angular momentum multiquarks may increase the stability of the system \cite{Karliner:2003dt,Bicudo:2010mv}.
The multiquark hamiltonians are important to understand not only the elusive multiquark hadrons, but also high density QCD where many quarks may overlap.

Experimentally, multiquark exotic hadrons have been searched for many years because as soon as the quarks were proposed in the sixties to classify the meson and baryon resonances, and the quark model was proposed in the seventies
\cite{DeRujula:1975ge}, 
it became clear that systems with more than three quarks could also possibly exist. 
One of the main problems of hadronic physics is thus to determine whether multiquark resonances exist or not, and weather the possible multiquark resonances are narrow or wide.

\begin{figure}[!t]
\begin{centering}
    \includegraphics[width=7cm]{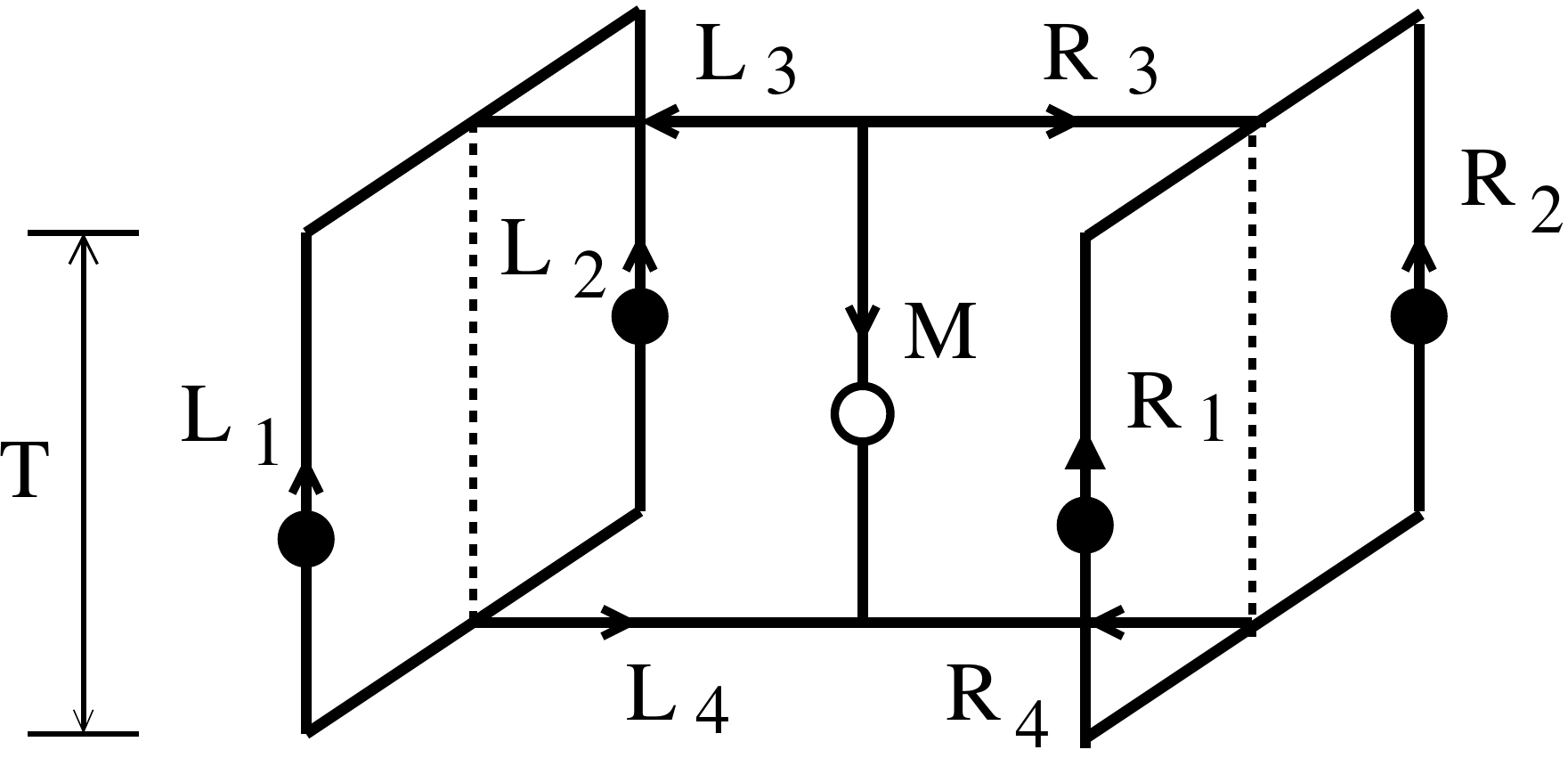}
\par\end{centering}
    \caption{Pentaquark Wilson loop as defined by Okiharu et al. \cite{Okiharu:2004wy,Okiharu:2005eg}. Here we extend this Wilson loop with different paths $L_i$ and $R_i$ for the quarks and $M_i$ for the antiquark.}
    \label{fig:pq_wl}
\end{figure}

The simplest multiquark system is the tetraquark, and it  was already  proposed by Jaffe in the seventies
\cite{Jaffe:1976ig} 
as a bound state formed by two quarks and two antiquarks. Presently some observed resonances are tetraquark candidates.
Very recently the Belle Collaboration made the tantalizing observation
\cite{Collaboration:2011gja},
in five different $\Upsilon$(5S)  decay channels 
of two new charged bottomonium resonances Z$_b$
with masses of 10610MeV/c$^2$ and 10650MeV/c$^2$ and narrow widths of the order or 15 MeV.  Since all standard bottomonia are neutrally charged, these two new resonances have a  flavour only compatible with $b \, \bar b  \, u  \, \bar d$ tetraquarks.
In 2003, the X(3872) observed by the Bell Collaboration  \cite{Choi:2003ue,Acosta:2003zx} was suggested as a tetraquark candidate by Maiani et al \cite{Maiani:2004vq}.
In 2004, the D$_{\text{sJ}}$(2632) state seen in Fermilab's SELEX \cite{Jun:2004wn,Cooper:2005zu} was suggested as a possible tetraquark candidate.
In 2009, Fermilab announced the discovery of Y(4140), which may also be a tetraquark \cite{Mahajan:2009pj}.
There are as well indications that the Y(4660) could be a tetraquark state \cite{Cotugno:2009ys}.
The $\Upsilon$(5S)  bottomonium has also been recently suggested to be a tetraquark resonance \cite{Ali:2009es}.
However a better understanding of tetraquarks is necessary to confirm or disprove the  X, Y, Z and possibly also other light resonances candidates as tetraquark states. 

The pentaquark is the next in the multiquark hadron series. Pentaquark hadrons were already proposed in the eighties by Manohar 
\cite{Manohar:1984ys}
and Chemtob
\cite{Chemtob:1985ar},
inspired by extensions of the Skyrme model.
In the 2000s a burst of interest was sparkled by a discovery claim of the $\Theta$ pentaquark by Nakano et al
\cite{Nakano:2003qx}.
This led to many experimental and lattice QCD studies of pentaquarks, together with hundreds of theoretical estimations of the $\Theta$ properties. However the resonance $\Theta$ ended up by not being confirmed by the scientific community
\cite{Dzierba:2004db,MartinezTorres:2010zzb}.
The many hundreds of publications on the subject, with disparate conclusions, show that the $\Theta$ pentaquark was beyond the scope of the scientific techniques utilized in the 2000s.

The multiquark hadrons are thus very elusive systems, much harder to observe experimentally, to understand in models, and to simulate in lattice QCD  than the conventional mesons and baryons. Nevertheless, inasmuch as the understanding of confinement, the existence/non-existence  of multiquark hadrons remain an important problem in QCD, to be further explored in the future PANDA experiment at GSI.

It is thus important to proceed with the well defined  program of understanding the static potentials and flux tubes of multiquarks in quenched Lattice QCD.

\begin{figure}[!t]
\begin{centering}
    \subfloat[\label{fig:geom10}]{
\begin{centering}
    \includegraphics[width=0.46\columnwidth]{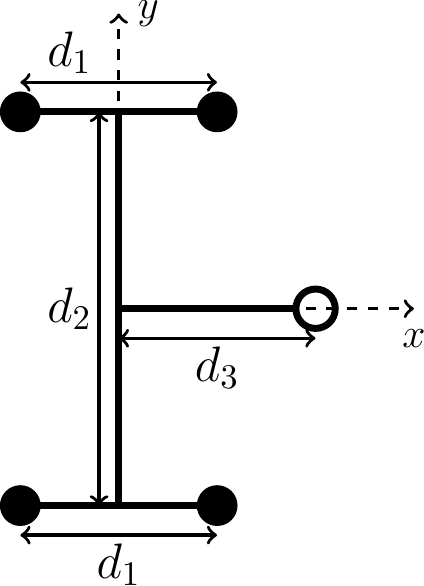}
\par\end{centering}}
    \subfloat[\label{fig:geom5}]{
\begin{centering}
    \includegraphics[width=0.46\columnwidth]{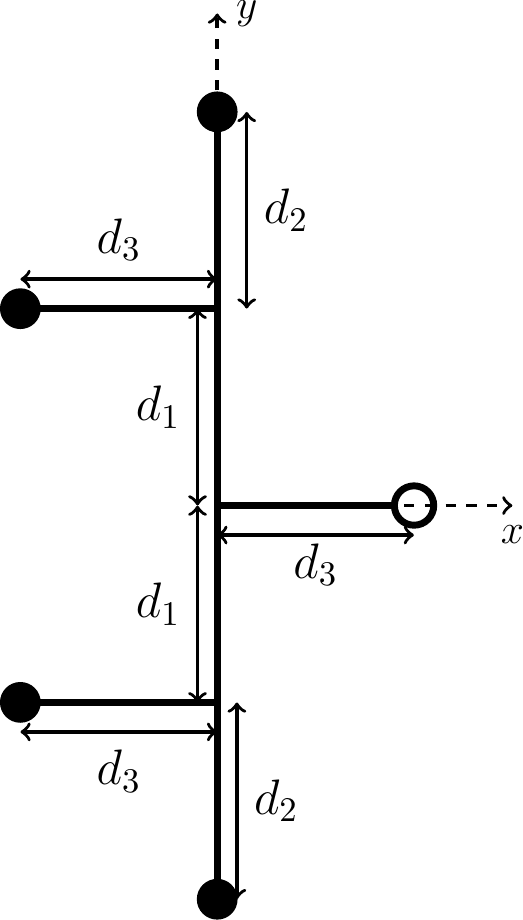}
\par\end{centering}}
\par\end{centering}
    \caption{Projection in the spacial dimensions of the different Wilson loop geometries for the static pentaquark studied in this work. The solid dots correspond to the quarks positions and the open dots to the antiquarks. The solid lines correspond to the space-like Wilson paths.}
    \label{fig:pq_geom}
\end{figure}

In the last years, the static tetraquark potential has been studied in Lattice QCD computations \cite{Alexandrou:2004ak,Okiharu:2004ve,Bornyakov:2005kn}. 
The authors concluded that when the quark-quark are well separated from the antiquark-antiquark, 
the tetraquark potential is consistent with One Gluon Exchange Coulomb potentials plus a four-body confining potential, 
suggesting the formation of a double-Y flux tube, typical of the four-body potential of the string flip-flop model as in Fig. \ref{fig:tq},
composed of five linear fundamental flux tubes meeting in two Fermat-Steiner points \cite{Vijande:2007ix,Bicudo:2008yr,Richard:2009jv}. 
This flux tube geometry was confirmed by Lattice QCD studies of the flux tubes produced by a static tetraquark system
\cite{Cardoso:2011fq,Cardoso:2012uk}.
In what concerns the pentaquark, static potentials have already been explored in a geometry with the antiquark situated in the centre of the four quarks
\cite{Okiharu:2004wy,Alexandrou:2004ak},
also consistent with a string flip-flop model, in this case with only six fundamental flux tubes and two Fermat-Steiner points.

\begin{table}[!t]
\caption{Pentaquark geometries studied and number of lattice configurations used in this work. The geometry type is outlined in Fig. \ref{fig:pq_geom}. The column \textbf{Id} corresponds to the numbering used in the text.}
\label{tab:pq_geom_details}
\begin{ruledtabular}
\begin{tabular}{cccccc}
\T\B \textbf{Id} & \textbf{Geometry Type}	& $\mathbf{d_1}$ & $\mathbf{d_2}$ & $\mathbf{d_3}$ & \# Configs.\\
\hline
\T\B (i) & Fig. \protect\ref{fig:geom10}	& 8 & 8 & 0 & 551\\
\T\B (ii) & Fig. \protect\ref{fig:geom10} 	& 8 & 8 & 6 & 549 \\
\T\B (iii) & Fig. \protect\ref{fig:geom5} 	& 4 & 4 & 8 & 544 \\
\T\B (iv) & Fig. \protect\ref{fig:geom5} 	& 6 & 4 & 8 & 1121 \\
\end{tabular}
\end{ruledtabular}
\end{table}

Here we proceed with the flux tube research program, studying the flux tubes of static pentaquarks in pure gauge SU(3) lattice QCD. In Section II we detail  the framework we set to measure the flux tubes.  We also extend the geometries explored in the static potential studies. In Section III we expose our results and conclude.

\section{Simulating the pentaquark flux tubes in lattice QCD}

\begin{figure*}[!t]
\begin{center}
\ifdefined\inclfigs
\begin{tabular}{ccccc}
&$\Braket{E^2}$ & $-\Braket{B^2}$ & $\mathcal{L}$ & $\mathcal{H}$\\
\begin{sideways}\hspace{1cm}Id (i)\end{sideways} & \includegraphics[width=4.2cm]{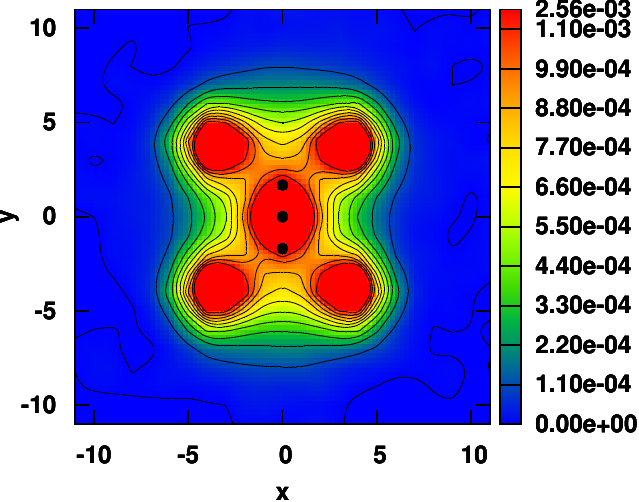} & \includegraphics[width=4.2cm]{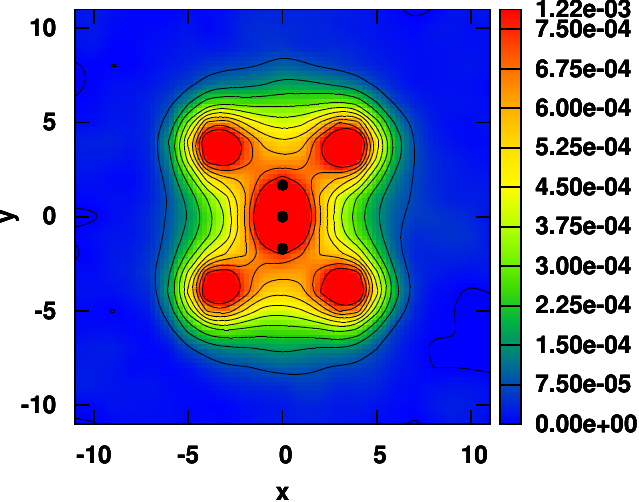} & \includegraphics[width=4.2cm]{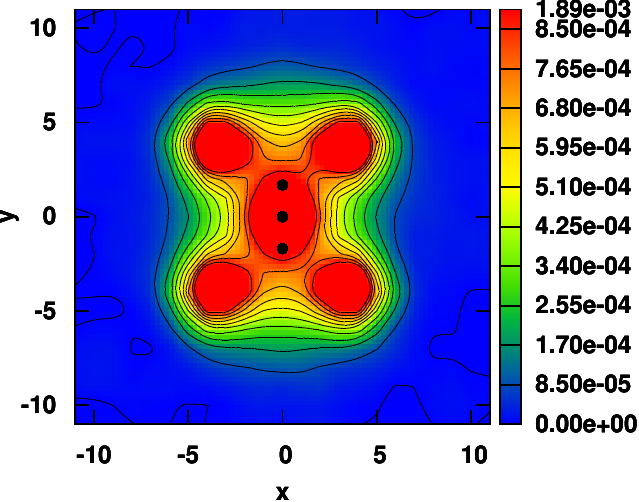} & \includegraphics[width=4.2cm]{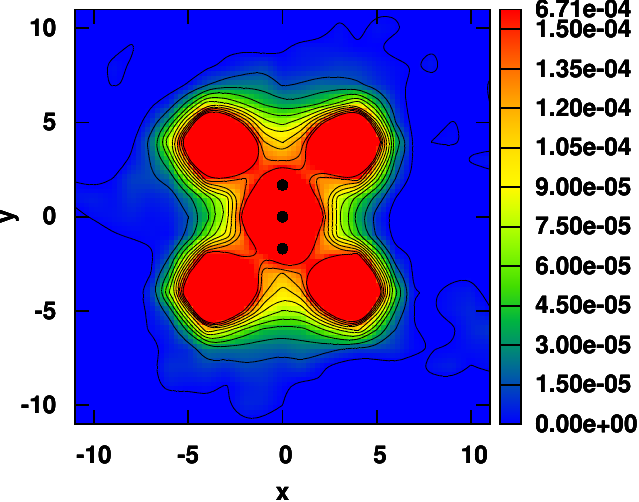}\\
\begin{sideways}\hspace{1cm}Id (ii)\end{sideways} & \includegraphics[width=4.2cm]{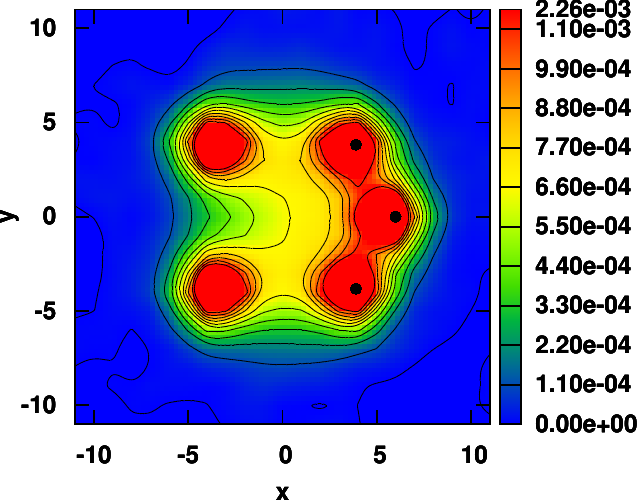} & \includegraphics[width=4.2cm]{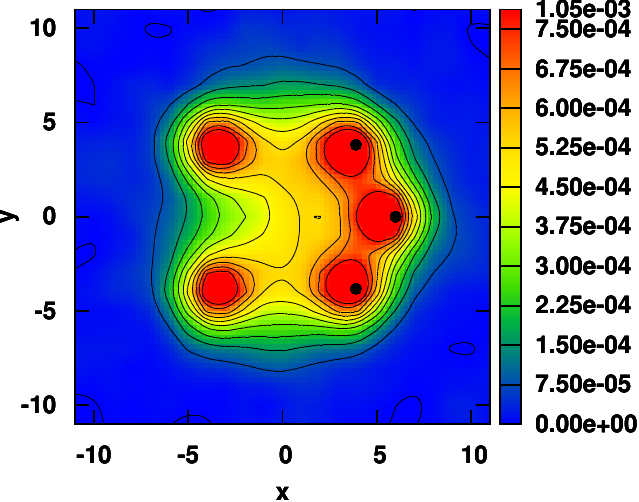} & \includegraphics[width=4.2cm]{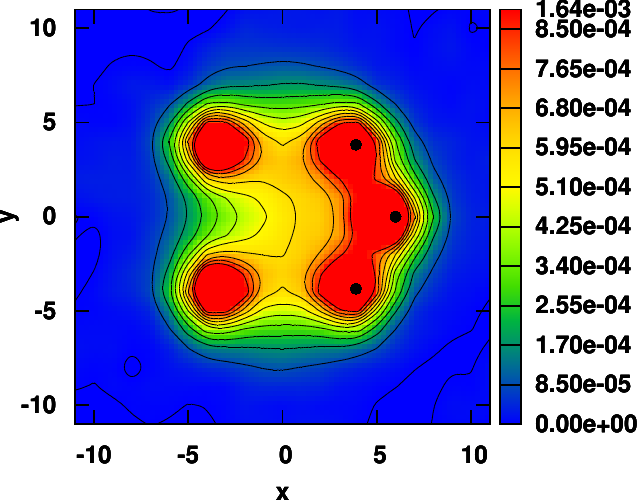} & \includegraphics[width=4.2cm]{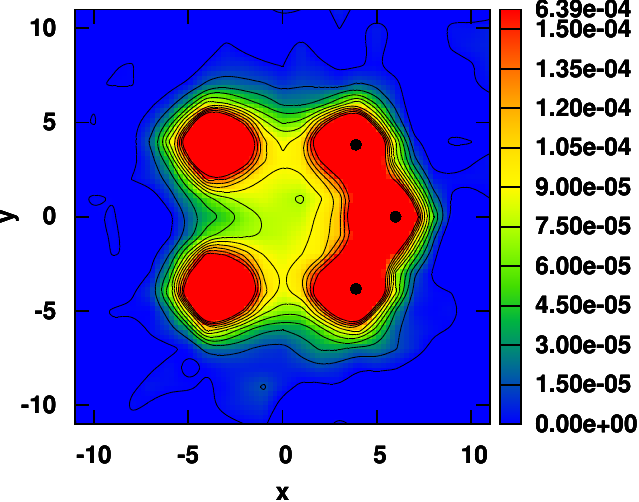}\\
\begin{sideways}\hspace{1cm}Id (iii)\end{sideways} & \includegraphics[width=4.2cm]{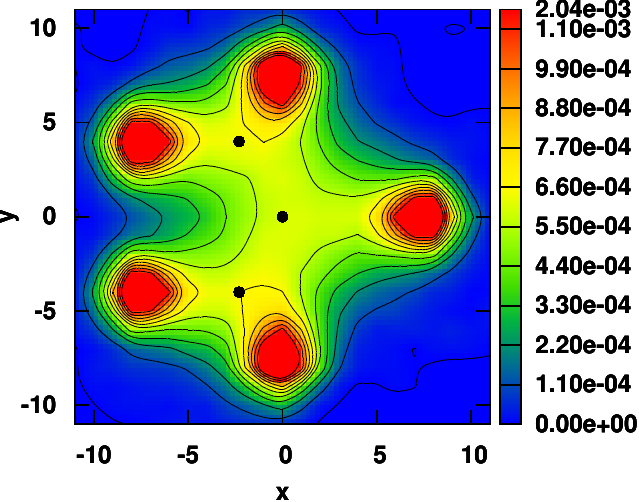} & \includegraphics[width=4.2cm]{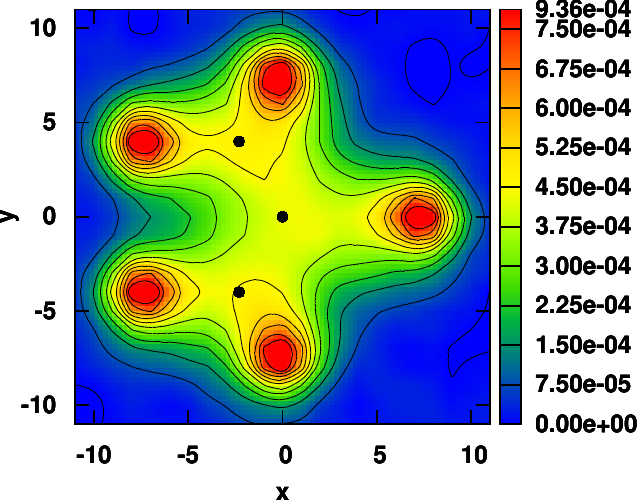} & \includegraphics[width=4.2cm]{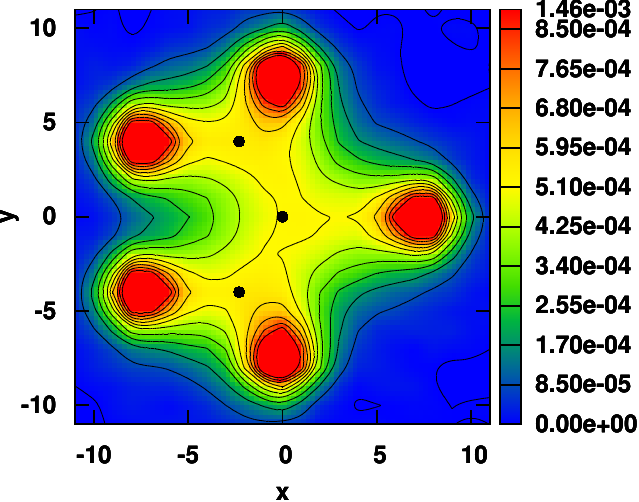} & \includegraphics[width=4.2cm]{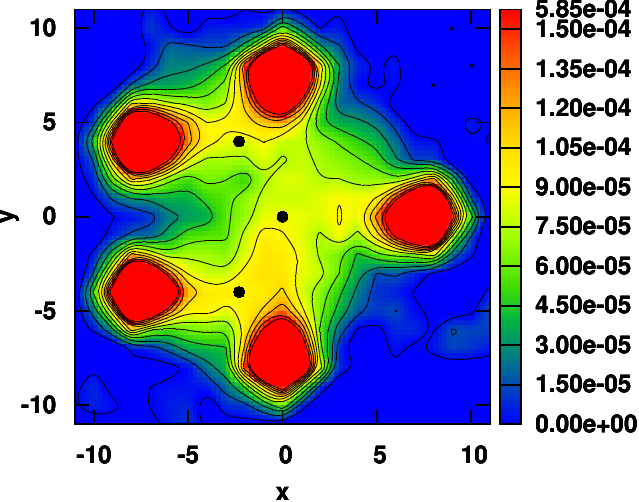}\\
\begin{sideways}\hspace{1cm}Id (iv)\end{sideways} & \includegraphics[width=4.2cm]{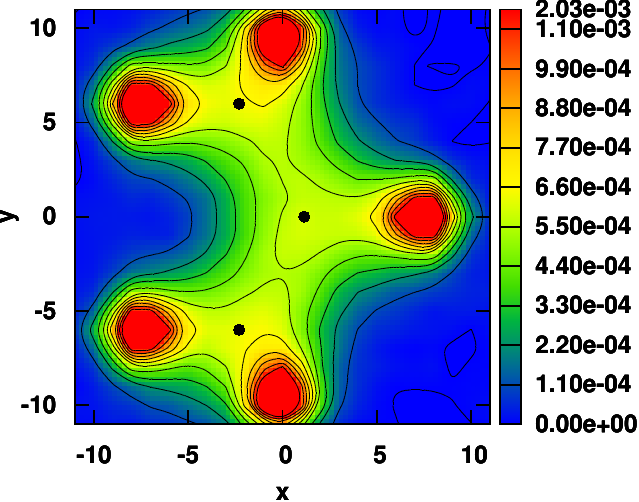} & \includegraphics[width=4.2cm]{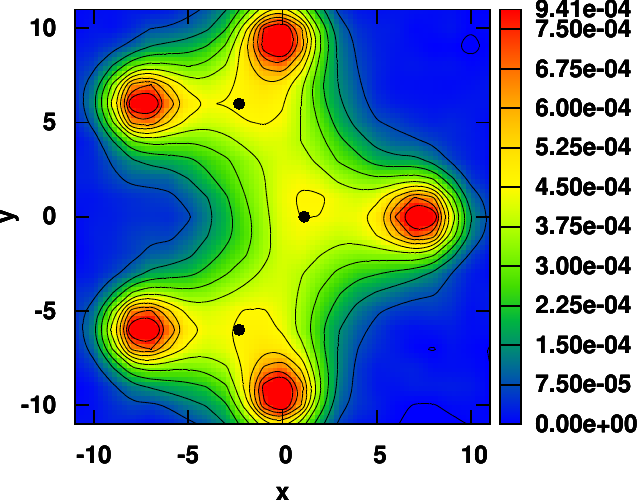} & \includegraphics[width=4.2cm]{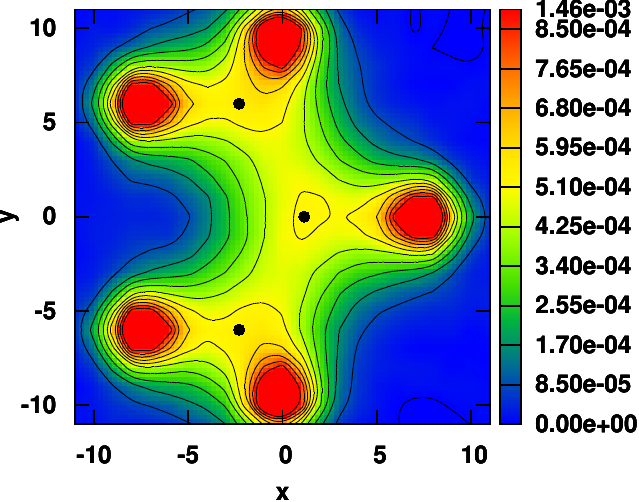} & \includegraphics[width=4.2cm]{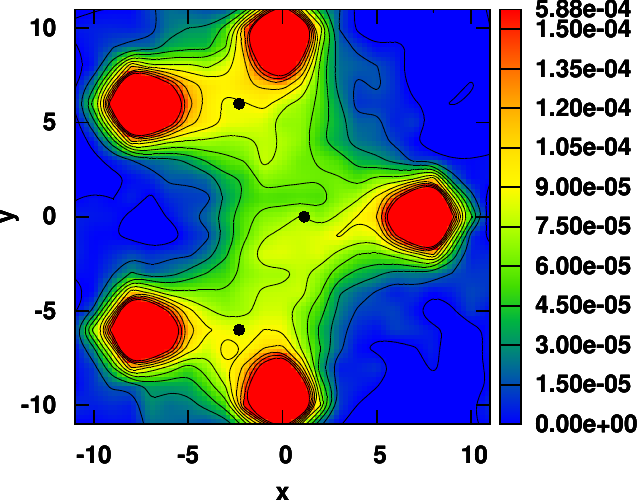}\\
\end{tabular}
\else\fi
    \caption{(Colour online.) Density plots of the chromoelectric and chromomagnetic fields and Lagrangian and energy densities for the geometries defined in Table \ref{tab:pq_geom_details}. The black dot points correspond to the Fermat-Steiner points, Table \ref{tab:pq_steiner_points}. The results are presented in lattice spacing units.}
    \label{fig:PQ_2D}
\end{center}
\end{figure*}

The static potential for the pentaquark was already studied in the lattice QCD by \cite{Alexandrou:2004ak} and \cite{Okiharu:2004wy,Okiharu:2005eg}
utilizing generalized Wilson loops. Here we use similar Wilson loops to place a static system of four quarks and one antiquark in the lattice, in four different geometries. Moreover we measure the colour-electric and colour-magnetic fields produced by the static charges.

\begin{figure*}[t!]
\ifdefined\inclfigs
\begin{centering}
    \subfloat[Id (i)\label{fig:PQ_Act_3D_10}]{
\begin{centering}
    \includegraphics[width=7cm]{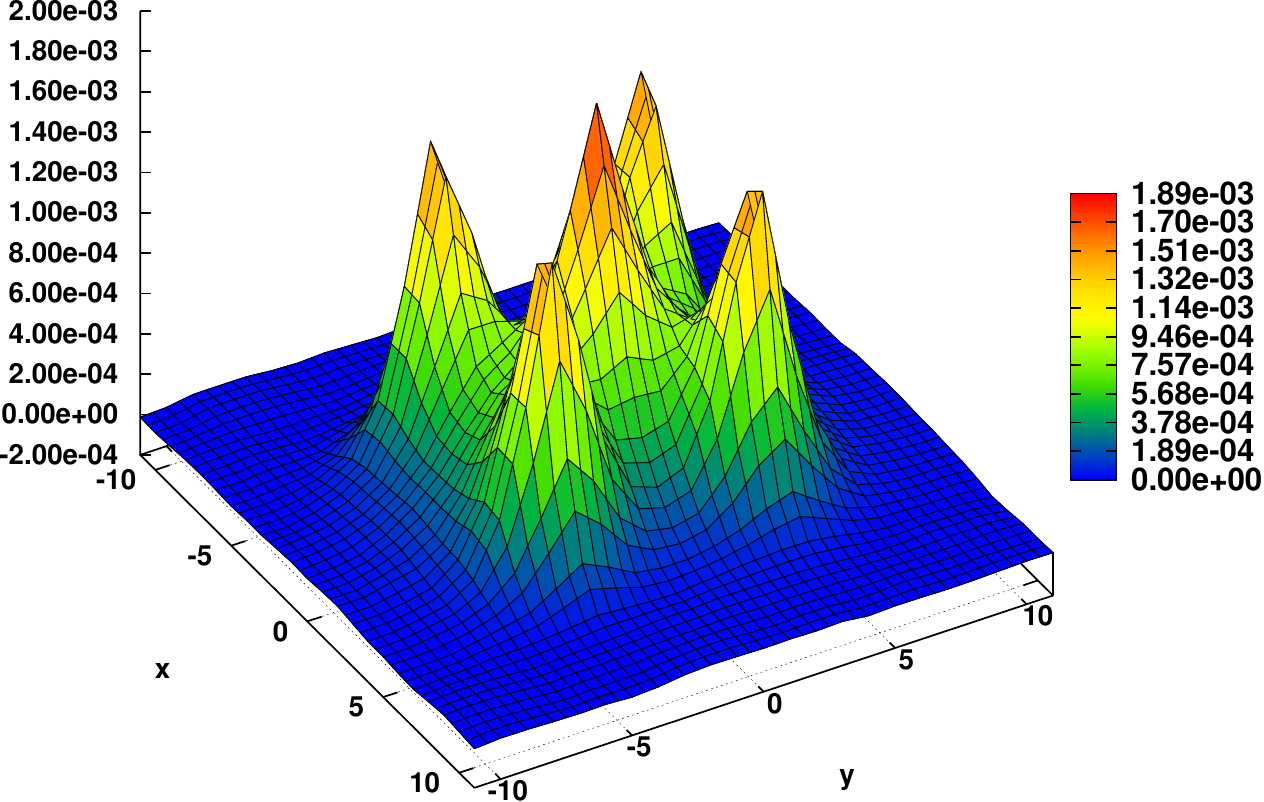}
\par\end{centering}}
    \subfloat[Id (ii)\label{fig:PQ_Act_3D_20}]{
\begin{centering}
    \includegraphics[width=7cm]{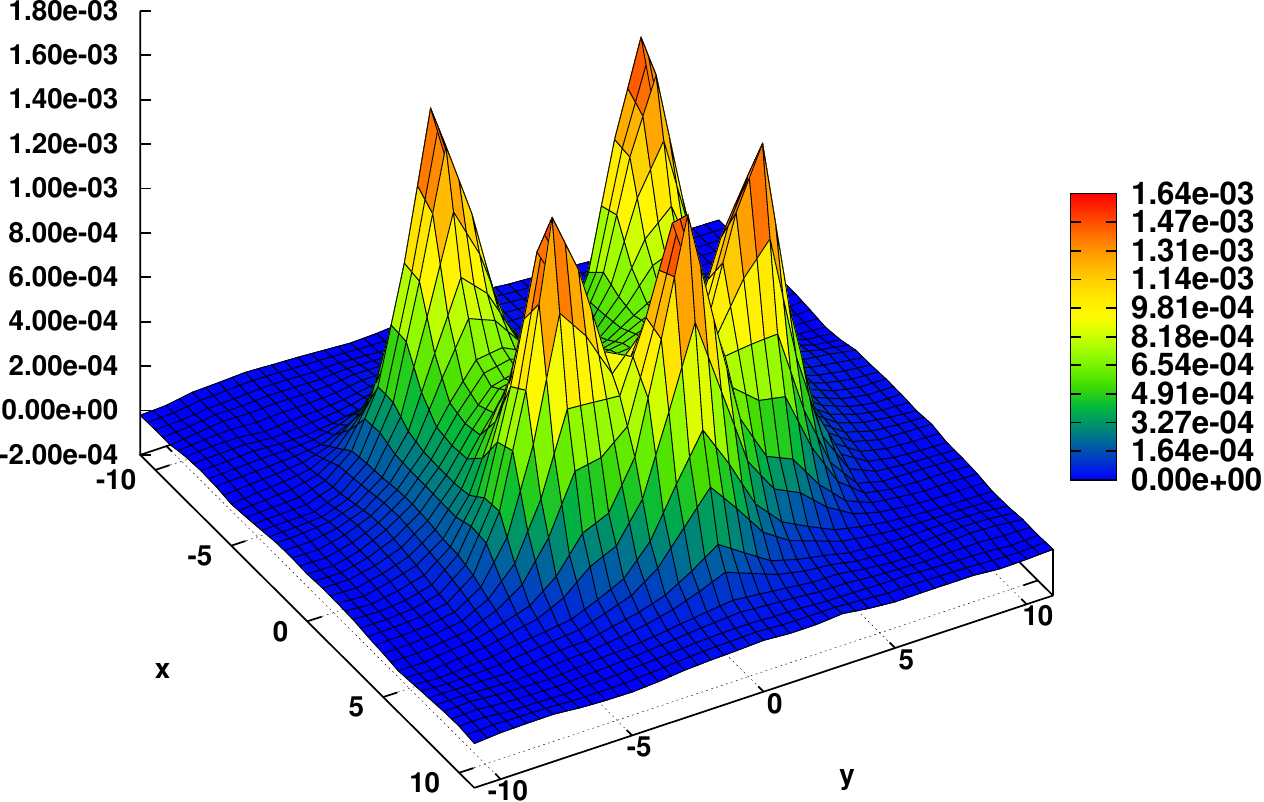}
\par\end{centering}}

    \subfloat[Id (iii)\label{fig:PQ_Act_3D_6}]{
\begin{centering}
    \includegraphics[width=7cm]{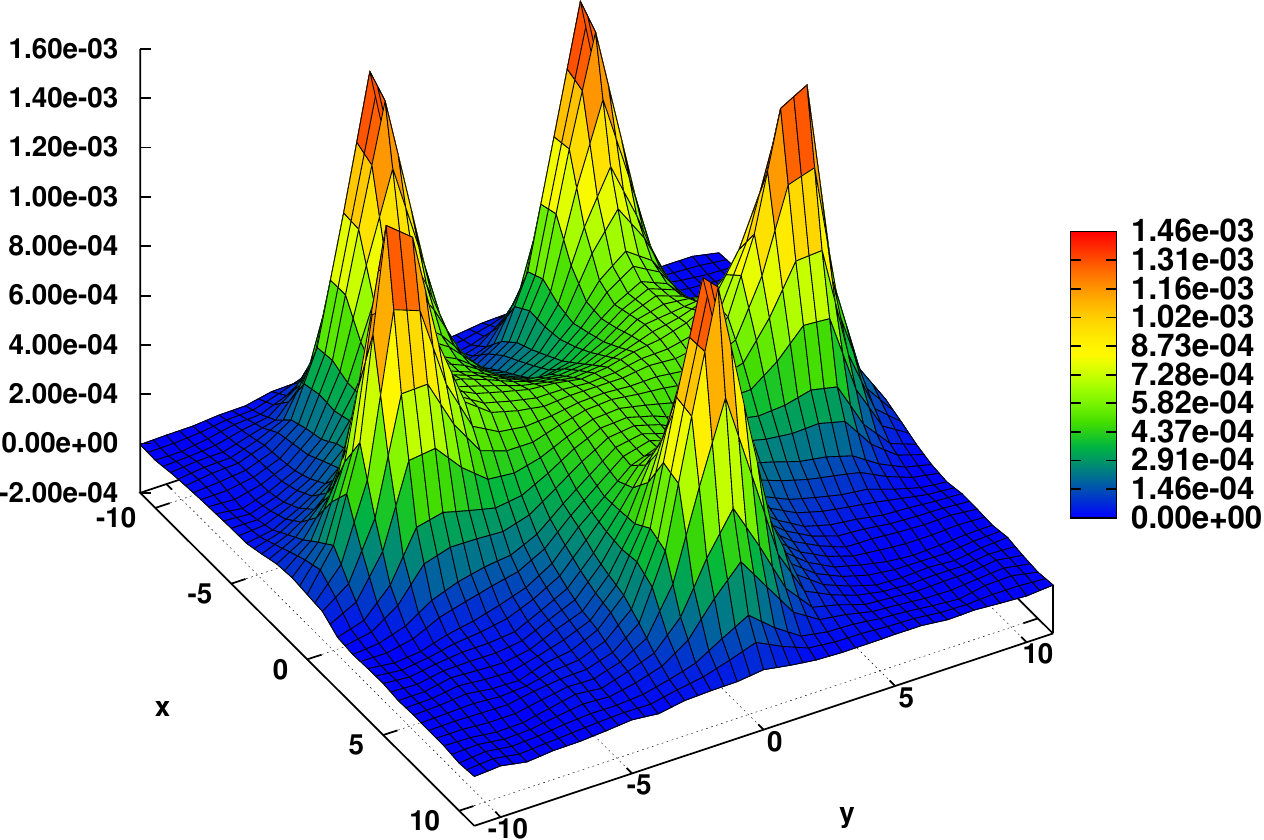}
\par\end{centering}}
    \subfloat[Id (iv)\label{fig:PQ_Act_3D_5}]{
\begin{centering}
    \includegraphics[width=7cm]{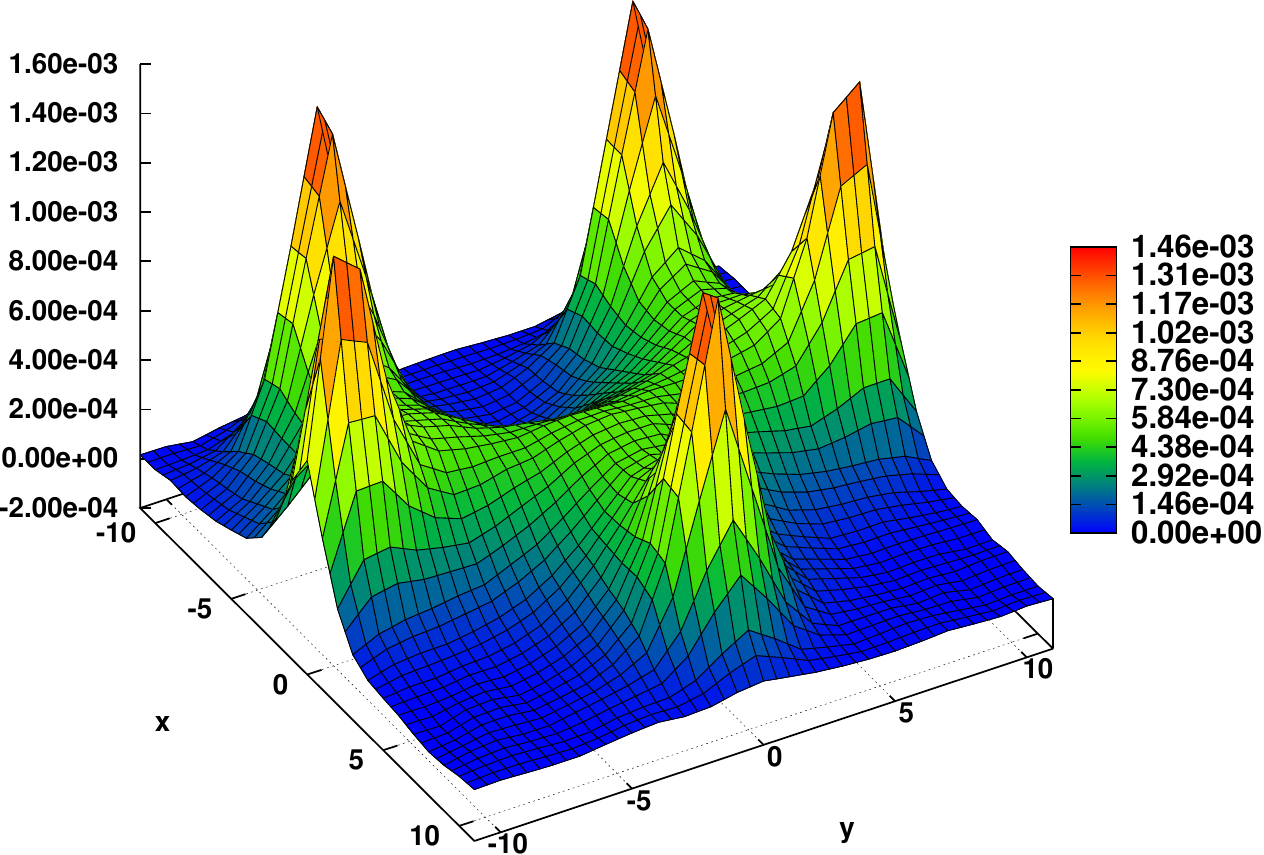}
\par\end{centering}}
\par\end{centering}
\else\fi
    \caption{(Colour online) we show three-dimensional plots of the Lagrangian density for the geometries defined in Table \ref{tab:pq_geom_details}. The density enhancement is maximal in the location of the colour charges, and if the colour charges were close the Coulomb potential would be important. With our geometries the colour charges are separated, and the fundamental flux tubes connecting the charges and the Fermat-Steiner points are evident.   The results are presented in lattice spacing units.}
    \label{fig:PQ_Act_3D}
\end{figure*}

The Wilson loop operator for the pentaquark system is defined in a gauge-invariant way, as illustrated in Fig. \ref{fig:pq_wl}, by
\begin{equation}
	W_{5Q} = \frac{1}{3!} \epsilon^{ijk}\epsilon^{i'j'k'} M^{ii'}\left(R_3 R_{12} R_4\right)^{jj'} \left(L_3 L_{12} L_4\right)^{kk'}\, ,
\end{equation}
where
\begin{eqnarray}
R_{12}^{i'i} &=& \frac{1}{2}\epsilon^{ijk}\epsilon^{i'j'k'}R_1^{jj'}R_2^{kk'}\,,\\ \nonumber
L_{12}^{i'i} &=& \frac{1}{2}\epsilon^{ijk}\epsilon^{i'j'k'}L_1^{jj'}L_2^{kk'} \,.
\label{Wilson path}
\end{eqnarray}
The projection in the spacial dimensions of our four different Wilson loop geometries for the static pentaquark is illustrated in  Fig.  \ref{fig:pq_geom}.
The distances of the geometries are quantified in Table \ref{tab:pq_geom_details}. The labelling of the geometries together with the number of lattice configurations used in this work are also shown in Table \ref{tab:pq_geom_details}.

\begin{table}[t]
\caption{Fermat-Steiner points for the pentaquark geometries studied. The geometry type is outlined in Fig. \ref{fig:pq_geom} and Table \ref{tab:pq_geom_details}.}
\label{tab:pq_steiner_points}
\begin{ruledtabular}
\begin{tabular}{ccccc}
\T\B & \multicolumn{3}{c}{\T\B \textbf{Fermat-Steiner Points}}\tabularnewline
\T\B \textbf{Id}	& $\mathbf{r_I}$ & $\mathbf{r_{II}}$ & $\mathbf{r_{III}}$\\
\hline
\T\B (i)	& $(0,-1.691,0)$ & $(0,1.691,0)$ & $(0,0,0)$\\
\T\B (ii) & $(3.897,-3.830,0)$ & $(3.897,3.830,0)$ & $(6,0,0)$ \\
\T\B (iii) & $(-2.309,-4,0)$ & $(-2.309,4,0)$ & $(0,0,0)$ \\
\T\B (iv) & $(-2.309,-6,0)$ & $(-2.309,6,0)$ & $(1.155,0,0)$ \\
\end{tabular}
\end{ruledtabular}
\end{table}

We compute the colour electric and the colour magnetic fields, by using the correlators of
the plaquettes $P_{\mu\nu}$ and the Wilson loop $W_{5Q}$.
We define the plaquettes as
$P_{\mu\nu} = 1 - \frac{1}{3} \mbox{Tr}[ U_\mu(\mathbf{s}) U_\nu(\mathbf{s}+\boldsymbol{\mu})
 U_\mu^\dagger(\mathbf{s}+\boldsymbol{\nu}) U_\nu^\dagger(\mathbf{s}) ]$.

With this definition, the chromofields are given by
\begin{eqnarray}
    \Braket{E^2_i} &=& \Braket{P_{0i}}-\frac{\Braket{W_{5Q}\,P_{0i}}}{\Braket{W_{5Q}}} \\
    \Braket{B^2_i} &=& \frac{\Braket{W_{5Q}\,P_{jk}}}{\Braket{W_{5Q}}}-\Braket{P_{jk}} \, ,
\end{eqnarray}
with the indices $j$ and $k$ complementing index $i$. 
The lagrangian and energy densities are given by $\mathcal{L}=\frac{1}{2}(E^{2}-B^{2})$
and $\mathcal{H}=\frac{1}{2}(E^{2}+B^{2})$.

To compute the static field expectation value, we plot the expectation value $\Braket{E^2_i(\mbf r)} $ or   $\Braket{B^2_i(\mbf r)}$ as a function of the temporal extension $T$ of the Wilson loop. 
At sufficiently large $T$, the groundstate corresponding to the studied quantum numbers dominates, and the expectation value tends to a horizontal plateau. 
In order to improve the signal to noise ratio of the Wilson loop, we use 50 iterations of APE Smearing with $w = 0.2$ (as in \cite{Cardoso:2009kz,Cardoso:2011fq}) in the spatial directions and one iteration of hypercubic blocking (HYP) in the temporal direction, \cite{Hasenfratz:2001hp}, with $\alpha_1 = 0.75$, $\alpha_2 = 0.6$ and $\alpha_3 = 0.3$.
Note that these two procedures are only applied to the Wilson Loop, not to the plaquette.

To check if the pentaquark flux tube produces a clear signal, we study the $\chi^2 /$dof of our pentaquark $T$ plateaux. 
But, surprisingly, event at some of the distances  
illustrated in Fig. \ref{fig:PQ_2D}, where the string flip-flop potential would favour the meson-baryon flux tube, with a lower energy than the pentaquark flux tube, we find $T$ plateaux with a good  $\chi^2$ /dof. This shows that the mixing between the pentaquark flux tube and the meson-baryon flux tube is small, and it is possible to study clear pentaquark flux tubes even at relatively large diquark distances.

To compute the fields, we fit the horizontal plateaux obtained for each point $\mbf r$ 
determined by the plaquette position, but we consider $z=0$ for simplicity. 
We finally compute the error bars of the fields with the jackknife method. 

We compute the Fermat-Steiner points with the iterative method of Bicudo et al. \cite{Bicudo:2008yr}. We have five quarks(antiquarks) with the label $i$ and three Fermat-Steiner points with label $a=I,I\hspace{-1pt}I,I\hspace{-1pt}I\hspace{-1pt}I$,
\begin{eqnarray}
\mbf r_i&=&(x_i,y_i,z_i)\ ,
\nonumber \\
\mbf r_a&=&(x_a,y_a,z_a)\ ,
\nonumber \\
r_{ia}&=&\sqrt{(x_a-x_i)^2+(y_a-y_i)^2+(z_a-z_i)^2} \ . 
\end{eqnarray} 
To minimize the total length of the strings, 
\be 
d=r_{1 \, I} + r_{2 \, I} +r_{3 \, I\hspace{-1pt}I} +r_{4 \, I\hspace{-1pt}I}+ r_{\bar 5 \, I\hspace{-1pt}I\hspace{-1pt}I} +r_{I \, I\hspace{-1pt}I\hspace{-1pt}I}+ r_{I\hspace{-1pt}I \, I\hspace{-1pt}I\hspace{-1pt}I}\ ,
\ee
we only need to solve one non-linear vector equation per Fermat-Steiner point,
\begin{eqnarray}
\mbf r_I  &=& { {\mbf r_1 \over r_{1\,I} }+ {\mbf r_2 \over r_{2\,I} }+{\mbf r_{I\hspace{-1pt}I\hspace{-1pt}I} \over r_{{I\hspace{-1pt}I\hspace{-1pt}I}\,I} }\over
{1 \over r_{1\,I} }+ {1 \over r_{2\,I} }+{1 \over r_{{I\hspace{-1pt}I\hspace{-1pt}I}\,I} }} \ ,
\nonumber \\
\mbf r_{I\hspace{-1pt}I}  &=& { {\mbf r_3 \over r_{3\,{I\hspace{-1pt}I}} }+ {\mbf r_4 \over r_{4\,{I\hspace{-1pt}I}} }+{\mbf r_{I\hspace{-1pt}I\hspace{-1pt}I} \over r_{{I\hspace{-1pt}I\hspace{-1pt}I}\,{I\hspace{-1pt}I}} }\over
{1 \over r_{3\,{I\hspace{-1pt}I}} }+ {1 \over r_{4\,{I\hspace{-1pt}I}} }+{1 \over r_{{I\hspace{-1pt}I\hspace{-1pt}I}\,{I\hspace{-1pt}I}} }} \ ,
\nonumber \\
\mbf r_{I\hspace{-1pt}I\hspace{-1pt}I}  &=& { {\mbf r_I \over r_{I\,{I\hspace{-1pt}I\hspace{-1pt}I}} }+ {\mbf r_{I\hspace{-1pt}I} \over r_{{I\hspace{-1pt}I}\,{I\hspace{-1pt}I\hspace{-1pt}I}} }+{\mbf r_{\bar 5} \over r_{\bar 5\,{I\hspace{-1pt}I\hspace{-1pt}I}} }\over
{1 \over r_{I\,{I\hspace{-1pt}I\hspace{-1pt}I}} }+ {1 \over r_{{I\hspace{-1pt}I}\,{I\hspace{-1pt}I\hspace{-1pt}I}} }+{1 \over r_{\bar 5\,{I\hspace{-1pt}I\hspace{-1pt}I}} }} \ .
\label{eq:steiner}
\end{eqnarray}

\section{Results and conclusion}

We remark that the signal is clear only if the paths considered in the Wilson loop overlap the flux tube. Thus we consider geometries for the Wilson loop where the paths are just some lattice spacings distant from the expected string position in the string flip-flop model. We consider the four different Wilson loop geometries, detailed in Fig. 
\ref{fig:pq_geom}
and in Table 
\ref{tab:pq_geom_details}.
We only utilize planar geometries for the colour sources, in order to produce clearer pictures of the fields.
The results for the colour field densities are presented only for the $x \, y$ plane since the colour sources are in this plane and the results with $z\neq 0$ are less interesting for this study. Then with colour field densities as a function of $x$ and $y$ we produce density plots and three-dimensional plots.

To produce the results presented in this work , we utilize quenched configurations in a $24^3 \times 48$ lattice at $\beta = 6.2$. The number of configurations used is described in Table \ref{tab:pq_geom_details}.
We present our results in lattice spacing units of $a$, with $a=0.07261(85)$ fm or $a^{-1}=2718\,\pm\, 32$ MeV. 
We generate our configurations in NVIDIA GPUs of the FERMI series (480, 580 and Tesla C2070) with a SU(3) CUDA code
upgraded from our SU(2) combination of Cabibbo-Marinari
pseudoheatbath and over-relaxation algorithm \cite{Cardoso:2010di,Cardoso:2011xu,ptqcd}.
Our SU(3) updates involve three SU(2) subgroups, we work with 9 complex numbers,
and we reunitarize the matrix.

\begin{figure}[!t]
\begin{centering}
    \includegraphics[width=8cm]{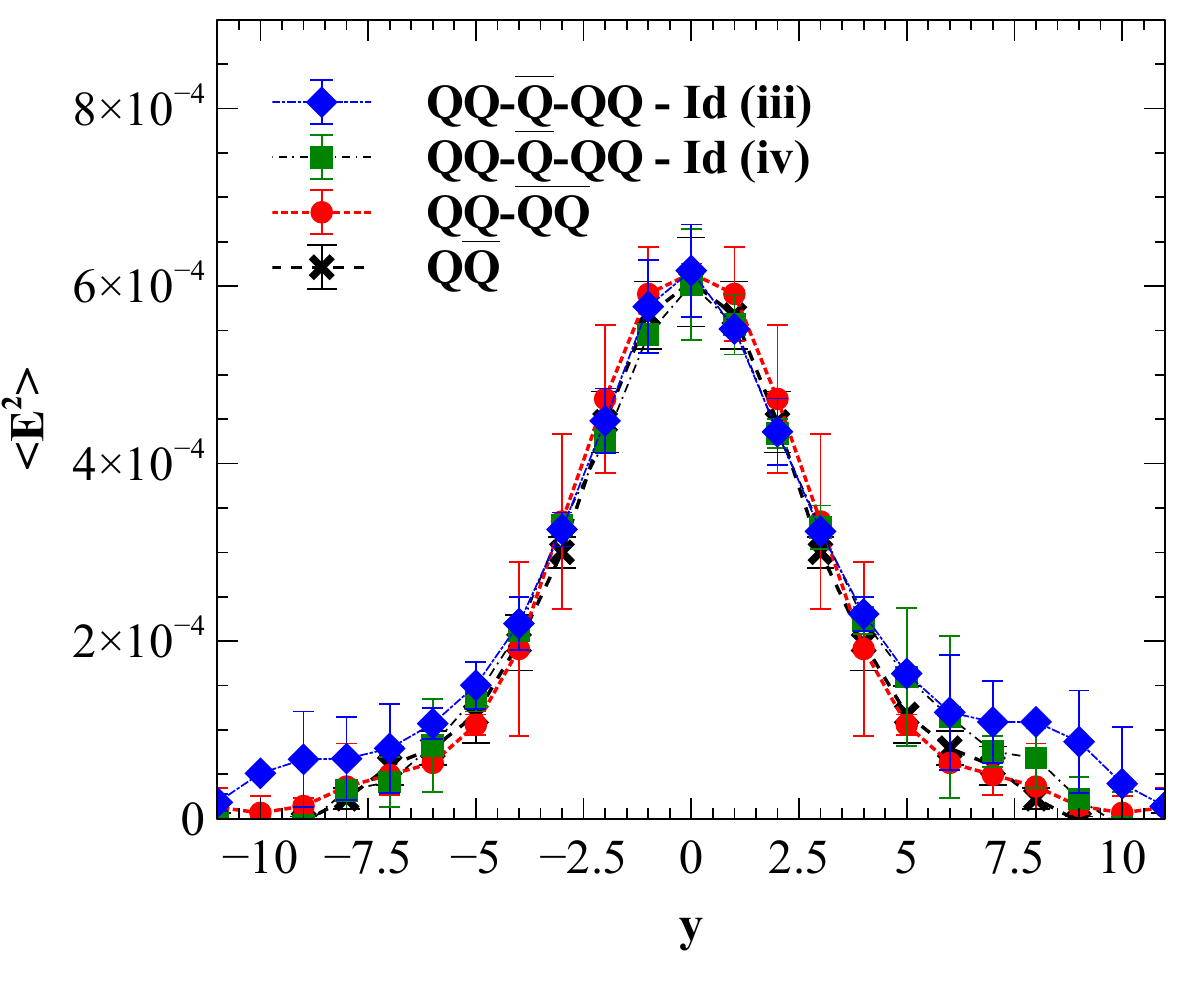}
\par\end{centering}
    \caption{(Colour online.) Profile of the chromoelectric field for the pentaquark, tetraquark and quark-antiquark systems. The pentaquark profile corresponds to the geometry  profile outlined in Fig. \ref{fig:pq_geom} and Table \ref{tab:pq_geom_details} along $x=4$.
    The tetraquark and quark-antiquark results are from \cite{Cardoso:2011fq}, in the middle of the flux tube.}
    \label{fig:pq_profile}
\end{figure}

The results for the colour fields, the energy and lagrangian densities are shown in Figs. \ref{fig:PQ_2D}$-$\ref{fig:PQ_Act_3D}. The figures clearly exhibit multi-Y-type shaped flux tubes. 
We also plot the Fermat-Steiner points defined in Table \ref{tab:pq_steiner_points}.
The Fermat-Steiner points of geometries (i) and (ii) are of different type from the Fermat-Steiner points of geometries (iii) and (iv), since in the first geometries angles of $120^{\circ}$ between the fundamental strings are not possible and thus the central Fermat-Steiner point has merged with the antiquark source. 
Nevertheless, and although the flux tubes have a finite width and are not infinitely thin as is assumed in the string flip-flop models, and although the Coulomb component of the potential is certainly important, we notice  the junctions for the elementary flux tubes are clearly close to the computed Fermat-Steiner points. This validates the use of string flip-flop models for the quark confinement in constituent quark models.

In Fig. \ref{fig:pq_profile}, we compare the chromoelectric field profile for the pentaquark, tetraquark and the quark-antiquark system in the middle of the flux tube. The tetraquark and the quark-antiquark results were obtained by \cite{Cardoso:2011fq}.
The three chromoelectric fields are identical up to the error bars. This confirms that the pentaquark flux tube is composed of a set of fundamental flux tubes with Fermat-Steiner junctions, and again validates the string flip-flop models as models for the quark confinement in constituent quark models.

Multiquark stability is a subtle theoretical problem, requiring the correct understanding and calibration of the quark interactions.
Combining our pentaquark results with the flux tube studies 
of mesons \cite{Bali:1994de}, 
baryons \cite{Takahashi:2004kc}, 
hybrids \cite{Cardoso:2009kz}, 
glueballs \cite{Cardoso:2008sb},
and tetraquarks \cite{Cardoso:2011fq,Cardoso:2012uk}
we finally feel confident that the string flip-flop potential, where fundamental strings with the minimal possible length link the static colour sources, is the correct phenomenological model for the confinement of any system of static quarks, antiquarks and gluons.
Whether the string flip-flop confining potential together with a correct short-range potential lead to multiquark narrow resonances or  boundstates remains a difficult quantum mechanical problem, but very interesting to the confinement and quark model experts.

\begin{acknowledgments}
This work was partly funded by the FCT contracts,  POCI/FP/81933/2007, 
CERN/FP/83582/2008, PTDC/FIS/100968/2008, CERN/FP/109327/2009, CERN/FP/116383/2010 and CERN/FP/123612/2011.
Nuno Cardoso is also supported by FCT under the contract SFRH/BD/44416/2008.
\end{acknowledgments}

\bibliographystyle{apsrev4-1}
\bibliography{bib_v1}

\end{document}